\documentstyle[12pt,epsfig]{article}

\def\eqn#1{(\ref{#1})}

\def\nn{\nonumber\\}

\def\ep{\varepsilon}
\def\o/{\"o}
\def\oo/{\H o}
\def\ii/{\'\i}
\def\u/{\"u}
\def\uu/{\H u}

\def\d{\partial}

\def\/{\hfill\break}
\def\be{\begin{equation}}
\def\ee{\end{equation}}
\def\ba{\begin{eqnarray}}
\def\ea{\end{eqnarray}}
\def\half{{1\over2}}

\def\exv#1{\left\langle #1\right\rangle}    

\def\delv{\rlap{\lower-.5ex\hbox{--}}{\delta}}

\def\cO#1{{\cal O}\left(#1\right)}

\def\p{{\bf p}}
\def\k{{\bf k}}

\def\tPhi{{\tilde\Phi}}
\def\tH{{\tilde H}}
\def\tG{{\tilde G}}

\def\fs{&&\hskip -0.6cm plus 0.1cm minus 0.1cm}

\def\DD{{\cal D}}

\begin{document}

\date{}
\title{
{\large\rm DESY 98-099}\hfill{\large\tt ISSN 0418-9833}\\
{\large\rm July 1998}\hfill\vspace*{2.5cm}\\
Viscosity of the Scalar Fields from the Classical Theory}
\author{A. Jakov\'ac\footnote{\em e-mail: jakovac@mail.desy.de}\\
{\normalsize\it Deutsches Elektronen-Synchrotron DESY, 22603 Hamburg, Germany}
\\[.2cm]
\vspace*{2cm}\\                     
}                                                                          

\maketitle  
\begin{abstract}
\noindent
We show how the resummation for time dependent quantities at high
temperature can be performed with an effective classical theory. As an
application we demonstrate that the leading term in the shear
viscosity, which is related to the $\rho_{\Phi^2\Phi^2}$ spectral
function can be calculated classically, either using classical linear
response theory or from the classical $\Phi^2$ correlation function. The
classical result depends explicitly on the cutoff, and the choice
$\Lambda\sim T$ reproduces the known quantum result.
\end{abstract} 
\thispagestyle{empty}
\newpage                                             


High temperature quantum field theories are known to suffer from a
number of IR problems which make the direct application of
perturbation theory unreliable. Part of these problems can be cured
by integrating out the hard thermal modes. For the static quantities
this yields dimensional reduction \cite{dimred}, for the nonstatic
ones, depending on the way we do it, we arrive at the HTL effective
action \cite{HTL} or an effective classical field theory
\cite{AS1}-\cite{BJ2}. In particular it was shown that in the $\Phi^4$
theory the self-energy can be calculated classically, and also the
first quantum correction can be reproduced from the effective theory
\cite{AS1,BJ1}.

The transport coefficients can be calculated from the microscopic
theory using linear response theory \cite{HST}. The shear viscosity
gets the dominant contribution from the spectral function of the
operator $\Phi^2(x)$ \cite{Jeon1}
\begin{equation}
  \label{eq:viscdef}
    \eta = \lim\limits_{p_0,\p\to0}\frac{\rho_{\Phi^2\Phi^2}(p_0,\p)}
    {p_0} = \lim\limits_{p_0,\p\to0}\,\frac{{\rm Disc}\,
    \exv{\,[\Phi^2(x),\Phi^2(0)]\,}}{p_0}.
\end{equation}
One can use a quantum theory with effective (resummed) spectral
functions to compute this quantity \cite{Jeon1,WHW,Jeon2}. The goal of
this paper is to show that the same results can be obtained from the
classical theory as well, which provides a simple calculational
possibility as well as a feasible numerical framework. 

In the followings we shortly recall (cf. Ref.s~\cite{AS2,NW,BJ2}) and
generalize, how one can use the classical theory to perform
resummation for time dependent quantities. We will then apply the
formalism to the viscosity.

\bigskip\noindent
{\Large\bf Resummation with Classical Theory}
\medskip

Dimensional reduction, which yields an extremely powerful method to
compute static quantities at high temperatures uses a very general,
renormalization group (RG) inspired technique to get rid of IR
divergencies: it identifies the most IR sensitive degrees of freedom,
separates them and integrates over the remaining ones. In this form it
can be generalized to develop a resummation method also for the
nonstatic quantities. To this end we have first to find the IR
sensitive degrees of freedoms, with other words the source of the IR
divergencies of the Feynman diagrams. The diagrams are generated by the
generating functional; for the $\Phi^4$ theory at finite temperature
it reads \cite{LaWe}
\begin{equation}
  Z[j]=e^{-i\int_c H_I(\frac\delta{i\delta j_\k(t)})}\, e^{-\half
  \int_c j_{-\k}(t) G(\k,t,t') j_\k(t')},
\end{equation}
where 
\begin{equation}
  G(\k,t)=\frac1{2\omega_\k}\left(e^{-i\omega_\k t}
  \Theta_c(t) + e^{i\omega_\k t} \Theta_c(-t)\right) + \frac
  {n(\omega_\k)} {\omega_\k} \cos\omega_\k t
\end{equation}
is the propagator, $n(\omega)=(e^{\beta\omega}-1)^{-1}$ the
Bose-Einstein distribution, $c$ is a real time contour (eg. the
Keldysh contour) and $\omega^2=\k^2+m^2$. In the IR regime where
$|\k|\ll T$ the vacuum part behaves as $\sim1/\omega$, the matter part
is $\sim T/\omega^2$, because $n(\omega)=T/\omega+\cO{1}$. For
massless fields the latter is quadratically, the former just linearly
diverges in the IR. Let us denote by $G_{IR}$ the IR sensitive part of
the propagator; it is defined asymptotically, its general form reads
\begin{equation}
\label{eq:IRprop}
  G_{IR}(\k,t)=\frac T{\omega_\k^2}\left(1 + \cO{\beta\omega}\right)\,
  \cos\omega_\k t.
\end{equation}
Let us moreover denote the IR regularized propagator by $\tG=G-G_{IR}$. 

The dimensional reduction technique suggests that we should rearrange
the perturbation theory and postpone the calculation with the most
singular propagator. We can try to represent the IR propagator by a
Gaussian path integral, like in the RG, and then interchange the order
of the operations. Symbolically, if
\begin{equation}
  \int\!d\phi \,e^{-\frac1{2 G_{IR}} \phi^*\phi+ ij\phi} \sim
  e^{-\half j G_{IR} j}
\end{equation}
then the effective theory is given by
\begin{equation}
  Z[j]=\int\!d\phi\,e^{-\frac1{2 G_{IR}} \phi^*\phi} \, e^{-\int
  H_I(\frac\delta{i\delta j })} \,e^{-\half j \tG j + ij\phi}.
\end{equation}
The perturbation theory generated by the interactions is now IR finite,
and provides an action for its background field which will be the
kernel of the subsequent path integral. In the standard RG we use the
formula to lower the cutoff by considering
$\tG(k)=\Theta(\Lambda>|k|>\Lambda-\Delta\Lambda) G(k)$, but in this
way we can introduce completely new type of degrees of freedom.

In our case the very special form of the time dependence of the IR
propagator allows {\em a 3D representation}. In fact we can write
\begin{equation}
  e^{-\half \int_c jG_{IR}j} = \int\DD\phi\DD\pi\,
  e^{-\tH_0[\phi,\pi] + i\int_c j \tPhi[\phi,\pi]},
\end{equation}
where 
\begin{eqnarray}
  \label{eq:ansatz}
  \fs \tH_0[\phi,\pi]=\int\!\frac{d^3\k}{(2\pi)^3}\,
  \frac1{2K_\k} \biggl( \pi_{-\k}
  \pi_\k\,+\,\omega_\k^2\,\phi_{-\k}\phi_\k \biggr)\nn 
  \fs \tPhi(\phi,\pi)=\int\!\frac{d^3\k}{(2\pi)^3}\,\biggl( \zeta_\k\,
  \phi_\k \,+\, \frac {\eta_\k}{\omega_\k}\,\pi_\k\biggr),
\end{eqnarray}
and, in the leading order
\begin{equation}
  \label{eq:leadchoice}
  K_\k=T,\quad \zeta_\k(t)=\cos \omega_\k t,\quad \eta_\k(t)=\sin
  \omega_\k t.
\end{equation}
Since the IR propagator is defined only in the leading order (cf.
\eqn{eq:IRprop}), we can freely choose the subleading parts of
$K,\,\zeta$ and $\eta$. The different choices of the subleading parts
lead to different 3D theories, corresponding to the different
prescriptions of the papers \cite{AS2,NW,BJ2}. We will use here
\begin{equation}
  \label{eq:ourchoice}
  K_\k= T\,\Theta(\Lambda-\omega_\k)\qquad
  \bigl(\zeta_\k(t),\eta_\k(t)\bigl)=\left\{
  \begin{array}[c]{ll}
    \bigl(\cos(\omega_\k t),\sin(\omega_\k t)\bigr), & {\rm if}
    \, t\in C_{1,2}\\
    (1,0), & {\rm if}\, t\in C_3,
  \end{array}\right.
\end{equation}
with the cutoff $\Lambda\sim T$ which describes the validity range of
the classical approximation \cite{BJ2}. The cutoff is useful also to
suppress the spatial nonlocalities in the effective action
\cite{JaPa}, and to ensure the quantum decoherence \cite{BJ2}.

To simplify notations we can introduce the effective generating
functional
\begin{equation}
  \tilde Z[j]=\frac1{\cal N}\, e^{-i\int_c H_I(\frac\delta{i\delta
  j})} \, e^{-\frac12 \int_c\int_c  j \tG j}\, e^{i\int_c j
  \tPhi[\phi,\pi]},
\end{equation}
where the normalizing factor ${\cal N}$ assures $\tilde Z[0]=1$ and
thus cancels vacuum diagrams. It is background dependent, and we will
include it into the Hamiltonian
\begin{equation}
  Z[j]=\int\DD\phi\DD\pi\, e^{-\tH[\phi,\pi]}\,\tilde Z[j],
\end{equation}
where $\tH=\tH_0+\ln {\cal N}$. Because of the special form of the
background field on the Matsubara contour \eqn{eq:ourchoice} the
effective Hamiltonian can be computed \cite{NW}
\begin{equation}
  \tH=\Gamma_{dim.red}[\phi] + \frac\beta2 \int\!\frac{d^3\k}{(2\pi)^3}\,
  \pi_{-\k} \pi_\k,
\end{equation}
where $\Gamma_{dim.red}[\phi]$ is the effective action of the
dimensional reduction.

The background field, beeing free, can be easily generated by the
initial conditions
\begin{equation}
  i\tPhi(t')=-\int_c dt\,\tilde J(t)\tG(t,t'),
\end{equation}
where
\begin{equation}
\label{eq:tildeJ}
  \tilde J(t)=(\d\tPhi(t_0)+\tPhi(t_0)\d_t)\delta_c(t-t_0)
\end{equation}
is the current localized at the initial time. This yields
\begin{equation}
  -\half\int_c j \tG j + i\int_c j\Phi = -\half\int_c J \tG J +
  \half\int_c \tilde J \tG \tilde J,
\end{equation}
where $J=j+\tilde J$. The last term is current-independent, so it is
canceled by the normalization. What remains is that $\tilde
Z[j,\tPhi]= \tilde Z[J,0]$, the background dependence can be absorbed
into a redefined current.

We can also perform a Legendre transformation with respect to this
current. The resulting effective action $\tilde\Gamma[\varphi]$ is
background independent, it can be calculated using the ordinary real
time perturbation theory with the propagator $\tilde G$. The 1PI
vertex functions can be obtained by differentiating $\tilde\Gamma$
with respect to $\varphi$, and take it at the physical point
$\varphi_{phys}$ which corresponds $j=0$
\begin{equation}
  \Gamma^{(n)}(x_1,\dots x_n)=\frac{\delta\tilde\Gamma[\varphi]}
  {\delta \varphi(x_1)\dots \delta \varphi(x_n)}
  \biggr|_{\varphi=\varphi_{phys}}. 
\end{equation}
$j=0$ means $J=\tilde J$. The Dirac deltas in $\tilde J$ at the
initial time set the initial conditions for the time evolution.
Therefore
\begin{equation}
   \frac{\delta\tilde\Gamma}{\delta \varphi}\biggr|_{phys} = 0,\qquad 
   \left\{\begin{array}[c]{l}
       \varphi_{phys}(t_0)=\phi\cr
       \d_t\varphi_{phys}(t_0)=\pi.
   \end{array}\right.
\end{equation}
That is, while the averaging over the initial conditions corresponds
to the local effective action, the time evolution is governed by the
time dependent effective action. The difference comes from the
time-nonlocal loops \cite{BJ2}, a genuine quantum effect.

\bigskip\noindent
{\Large\bf Shear Viscosity in Scalar Field Theories}
\medskip

As an application we can calculate the shear viscosity in $\Phi^4$
theory. First we recall the quantum result \cite{Jeon1,WHW}, then the
different classical approaches.

\paragraph{The quantum result}

The calculation is not too involved even in the quantum case. The
retarded Green functions have to be computed by the rule
\begin{equation}
  G^R_{AB}(x)=G^{11}_{AB}(x) - G^{12}_{AB}(x).
\end{equation}
In the Feynman diagrams we have to use the propagators
\begin{eqnarray}
  \fs iG^{11}(k)=iG^R(k)+iG^{12}(k),\quad 
  iG^{12}(k)=n(k_0)\rho(k),\nn
  \fs iG^{22}(k)=-iG^R(k)+iG^{21}(k),\quad 
  iG^{21}(k)=(1+n(k_0))\rho(k),\nn
\end{eqnarray}
where $iG^R(\k,t) = \Theta(t)\,\rho(\k,t)$. We have performed here
a self-consistent resummation and used interacting spectral function
instead of the free one (cf.  \cite{Jeon1, WHW}). This yields
(including the symmetry factor 2)
\begin{equation}
  iG^R_{\Phi^2\Phi^2}(\p,t) = 2\int\frac{d^3\k}{(2\pi)^3}\,\Theta(t)\,
  \left[ \rho(\k,t) \rho(\p-\k,t) + 2 \rho(\k,t) iG^<(\k,t) \right].
\end{equation}
Performing a Fourier transformation in time
\begin{equation}
  G^R_{\Phi^2\Phi^2}(p) = 2\int\frac{d^3\k}{(2\pi)^3}
  \frac{d\omega}{2\pi} \frac{d\omega'}{2\pi}\, \frac{\rho(\k,\omega)
  \rho(\p-\k,\omega')}{ p_0-\omega-\omega'+i\ep}\,(1+n(\omega) +
  n(\omega')).
\end{equation}
The discontinuity can be simply obtained using ${\rm Disc}(x+i\ep)^{-1}
=- 2\pi i\delta(x)$
\begin{equation}
\label{eq:qres}
  \rho_{\Phi^2\Phi^2}(p) = 2\int\frac{d^4k}{(2\pi)^4} \, \rho(k)
  \rho(p-k)\, (1+n(k_0) + n(p_0-k_0)).
\end{equation}
Using the identity
\begin{equation}
  1+n(\omega) + n(\omega')= (1-e^{-\beta(\omega+\omega')})(1+n(\omega))
  (1+n(\omega')) 
\end{equation}
we get back the previous results \cite{Jeon1,WHW}.

\paragraph{The classical approach}

\begin{figure}[htbp]
  \begin{center}
    \epsfig{file=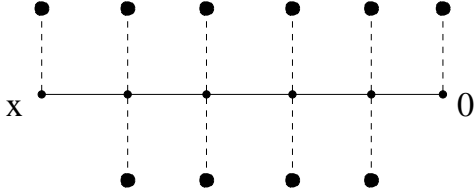,height=2.5cm}
    \caption{\em A typical classical contribution to the effective
      operator.}
    \label{fig:clcont2}
  \end{center}
\end{figure}

As it is proven before, the same quantum result can be obtained using
a two-step method, where in the first step we compute the effective
operator only, with the IR stable propagator and in the background of
the IR fields. A typical diagram contributing to the classical
(ie. without quantum loops) part for the retarded Greens function is
shown on Fig.~\ref{fig:clcont2}.
In formula
\begin{equation}
\label{eq:phi2eff}
  G^R_{\Phi^2\Phi^2,eff}(x)=4\Phi(x)\,G^R_{\Phi\Phi}(x,0;\Phi)\,\Phi(0),
\end{equation}
where $G^R_{\Phi\Phi}(x,0;\Phi)$ is the retarded Greens function
(cf. \cite{BJ1,BJ2}).

The same result can be obtained from the classical linear response
theory. To see it let us add a current term to an action
\begin{equation}
  S(j)=S + \int j f(\Phi),
\end{equation}
where $f$ is an arbitrary function. We want to examine the linear
response of the function $g(\Phi)$ to this change
\begin{equation}
  G^R_{fg}(x-x')= \frac{\delta g(\Phi(j,x))}{\delta j(x')}\biggr|_{j=0} =
  g'(\Phi(0,x)) G^R_{f\Phi}(x-x').
\end{equation}
We can get $G^R_{f\Phi}$ from the modified equations of motion
\begin{equation}
  0=\frac{\delta S(j)}{\delta\Phi(x)} = \frac{\delta S}{\delta\Phi(x)}
  + j(x) f'(\Phi(x))
\end{equation}
by differentiating with respect to $j$:
\begin{equation}
  \int \frac{\delta S}{\delta\Phi(x)\delta\Phi(x')} G^R_{f\Phi}(x'-y)
  = - f'(\Phi(y)) \,\delta(x-y).
\end{equation}
Its solution is simply
\begin{equation}
  G^R_{f\Phi}(x-y)= G^R_{\Phi\Phi}(x-y) \,f'(\Phi(y)),
\end{equation}
so finally
\begin{equation}
  G^R_{fg}(x-x')=g'(\Phi(x))\, G^R_{\Phi\Phi}(x-x')\, f'(\Phi(x')).
\end{equation}
For $f=g=\Phi^2$ we reproduce \eqn{eq:phi2eff}.
\newpage

After the 3D integration the background lines are closed with
\begin{equation}
  \exv{\Phi(k)\Phi(q)} = (2\pi)^4 \delta(k+q) \,iG_{3D}(k).
\end{equation}
The background fields, however, also follow a nontrivial, nonlinear
time evolution, and the complete calculation cannot be performed in
its generality. We will make an approximation similar to the one in
the quantum theory: we join all the background lines on one propagator
(i.e. sum up the the self-energy diagrams) and work further with these
effective propagators. This yields the spectral representations
\begin{equation}
\label{eq:effprop}
  G_R(k)=\int\frac{d\omega}{2\pi}\,\frac{\rho(\k,\omega)}
  {k_0-\omega+i\ep}, \qquad
  iG_{3D}(k)= \frac T{k_0}\,\rho(k),
\end{equation}
where we should use the complete spectral functions instead of the
free ones. The retarded Greens function reads
\begin{equation}
  iG^R_{\Phi^2\Phi^2}(p)= 4\int\frac{d^4k}{(2\pi)^4}\, iG_R(p-k)\,
  iG_{3D}(k),
\end{equation}
which yields the discontinuity (note the symmetry $k_0\leftrightarrow
p_0-k_0$)
\begin{equation}
\label{eq:clres}
  \rho_{\Phi^2\Phi^2}(p)=2T\int\frac{d^4k}{(2\pi)^4} \,
  \rho(p-k)\rho(k)\, \left(\frac1{k_0}+\frac1{p_0-k_0}\right).
\end{equation}

There is another use of the approximation \eqn{eq:effprop}, because it
provides a simple way to extract the spectral function (cf. \cite{WHW}
for the quantum case)
\begin{equation}
  \rho_{AB}(p) = \beta p_0\,iG_{AB,3D}(p).
\end{equation}
In our case we have to compute the classical expectation value
\begin{equation}
  \exv{\Phi^2(x)\Phi^2(0)}_{class} = 2iG_{3D}(x)\, iG_{3D}(x)
\end{equation}
and, after Fourier transformation
\begin{equation}
\label{eq:2pf}
  \rho_{\Phi^2\Phi^2}(p)= 2\beta p_0\int\frac{d^4k}{(2\pi)^4}\,
  \frac T{k_0}\rho(k)\,\frac T{p_0-k_0}\rho(p-k) 
\end{equation}
which is indeed identical with \eqn{eq:clres}.

Let us finally compute the viscosity from the classical theory. We
approximate the spectral function by
\begin{equation}
  \rho(k)\approx\frac{4k_0\gamma_\k}{(k_0^2-\omega_\k^2)^2+
  16k_0^2\gamma_\k^2},
\end{equation}
which yields in the narrow width approximation \cite{Jeon1,WHW}
\begin{equation}
\label{eq:narwidth}
  \rho(k)^2 \approx\frac{\pi\delta(k_0^2-\omega_\k^2)}
  {\omega_\k\gamma_\k}.
\end{equation}
To be consistent we have to use the classical value of the damping
rate. Its parametric form has been given by \cite{BJ1}, it is
equivalent to a high temperature approximation of the parametric
quantum result \cite{Jeon2,WH,Wel}. Therefore we can extract the on shell
classical damping rate from the corresponding quantum expression \cite{WH}
performing a high temperature expansion
\begin{equation}
\label{eq:clgam}
  \gamma_\k=\frac{\lambda^2T^2}{1536\pi\omega_\k}\left[1-\frac6{\pi^2}
  \int\limits_0^{|\k|}\!\frac{dq}{|\k|}\,L_2\left(1-\frac{\omega_\k^2}
  {\omega_q^2} \right)\right],
\end{equation}
where $L_2=-\int_0^z\ln(1-t)/t\,dt$ is the Spence function. After
interchanging the order of integrations we can write
\begin{equation}
  \int\limits_0^{|\k|}\!\frac{dq}{|\k|}\,L_2\left(1-\frac{\omega_\k^2}
  {\omega_q^2} \right) = - \int\limits_0^{k^2/m_T^2}\! dt\,
  \frac{\ln(1+t)}{t\sqrt{1+t}}\,\sqrt{1- \frac{tm_T^2}{k^2}}.
\end{equation}
The integral vanishes for $|\k|=0$; for large $|\k|$ we find
\begin{equation}
\label{eq:largemom}
  \gamma_\k\omega_\k\bigr|_{|\k|\gg m_T}=
  4\gamma_0\omega_0\,\left(1+\cO{\frac{m_T}{|\k|}\ln \frac{|\k|}{m_T}}
  \right).
\end{equation}
From \eqn{eq:viscdef}, \eqn{eq:2pf} and \eqn{eq:narwidth} we then
obtain
\begin{eqnarray}
  \eta\fs =\lim\limits_{p\to 0}\frac{\rho_{\Phi^2\Phi^2}(p)}{p_0} =
  2T\int\frac{d^4k}{(2\pi)^4}\, \frac{\rho(k)^2}{k_0^2}\approx
  T\int\frac{d^3\k}{(2\pi)^3}\, \frac 1 {\gamma_\k\omega_\k^4}=\nn
  \fs = \frac{768}{\lambda^2T\pi} \int\limits_0^\Lambda\! dk\,
  \frac{k^2}{(k^2+m_T^2)^{3/2}}\,\frac{\gamma_0\omega_0}
  {\gamma_\k\omega_\k}.
\end{eqnarray}
Since $\gamma_\k\omega_\k$ is bounded, $\eta$ will have a logarithmic
divergence, its coefficient can be extracted from the large momentum
behaviour of $\gamma_\k\omega_\k$ (cf. \eqn{eq:largemom})
\begin{equation}
  \eta= \frac {192}{\lambda^2T\pi}\left[\ln\frac\Lambda{m_T} \,+\,
  {\rm const.}\right].
\end{equation}
With $\Lambda\sim T$ we get back the leading term of the quantum
result \cite{WHW}. Using the complete expression \eqn{eq:clgam} we
find const$=2.1552$.

This divergence cannot be canceled by any local counterterm in the
Lagrangian. It is the consequence of having a composite operator which
needs renormalization even in the classical case. The complete result,
of course is independent on this auxiliary cutoff, the UV integration
should carry the appropriate counterterms.

\bigskip\noindent
{\Large\bf Conclusion and Outlook}
\medskip

We have summarized, how an effective classical theory can solve the
resummation problems in high temperature quantum field theories. For
static quantities this classical theory is equivalent to dimensional
reduction, the time evolution is governed by the effective quantum
action.

With this effective classical theory we could reproduce in the
$\Phi^4$ model the quantum result for the shear viscosity. It could be
computed either by using the definition (from the retarded Greens
function) or directly from the 3D expectation value of
$\exv{\Phi^2(x)\Phi^2(0)}_{class}$. The usual classical theory gives a
cutoff dependent result even after a proper renormalization. This
cutoff dependence has to vanish if we calculate also the UV
contributions to the effective operator.

The viscosity in this form is not complete, as shown in \cite{Jeon2}.
The correction terms (ladder diagrams) can be of the same order as the
leading one. Summing them up is far from beeing trivial. Since,
however, the important $1/\lambda^2$ behaviour of the viscosity is
essentially classical, one can try to perform the summation of the
ladder diagrams in this approach, which may be easier than in the
quantum theory. This is a task for future studies.

The classical theory, on the other hand, can be simulated on
computers. Since the viscosity is directly proportional to the
$\Phi^2$ two-point function, it is a relatively simply accessible
quantity. The simulations have to be performed with a finite cutoff of
the order of the temperature, or with the proper renormalization
factor stemming from the UV integration.

\medskip {\bf Acknowledgment}: I would like to thank U. Heinz and E.
Wang for calling my attention to this problem and for discussions, and
D. B\"odeker, W.  Buchm\"uller, Z. Fodor, A.  Patk\'os and P.
Petreczky for discussions. This work has been partially supported by
the Hungarian NFS under contract OTKA-T22929.


\end{document}